\documentclass[desactivate]{aa}
\newcommand{\ov}{\alpha_{\mathrm{ov}}} 
\newcommand{\conv}{\alpha_{\mathrm{conv}}} \newcommand{\dd}{\mathrm{d}}
\newcommand{\teff}{T_{\mathrm{eff}}} 
 
\newcommand{\smass}{M_{\odot}} 
\newcommand{\metal}{[\mathrm{Fe}/\mathrm{H}]} \newcommand{\gradad}{\nabla_{\rm
ad}} \newcommand{\gradrad}{\nabla_{\mathrm{rad}}}
\newcommand{\dov}{d_{\mathrm{ov}}}

\newcommand{\li}{\element[][7]{Li}} \newcommand{\be}{\element[][7]{Be}}
\newcommand{\heq}{\element[][4]{He}} \newcommand{\het}{\element[][3]{He}}
\newcommand{\h}{\element[][1]{H}} \newcommand{\deut}{\element[][2]{H}}
\newcommand{\bh}{\element[][8]{B}} \newcommand{\rcc}{R_\mathrm{cc}}

\usepackage{graphicx}
\usepackage{txfonts}
\usepackage{natbib}
\usepackage{hyperref}
\hypersetup{ colorlinks, linkcolor=blue, citecolor=blue }
\usepackage{float}

\begin{document}

\title{Impact of central mixing scheme and nuclear reaction network on the extent of convective cores}

\author{Anthony Noll\inst{1,2}
\and S\'ebastien Deheuvels\inst{2}}
\authorrunning{A. Noll \& S. Deheuvels}
\institute{Heidelberger Institut für Theoretische Studien, Schloss-Wolfsbrunnenweg 35, 69118 Heidelberg, Germany\label{1}
 \and IRAP, Université de Toulouse, CNRS, CNES, UPS, 14 avenue Edouard Belin, 31400 Toulouse, France\label{2}}

\offprints{A. Noll\\ \email{anthony.noll@h-its.org}} \abstract{Convective cores
are the hydrogen reservoirs of main sequence stars that are more massive than
around 1.2 solar masses. The characteristics of the cores have a strong impact
on the evolution and structure of the star. However, such results rely on
stellar evolution codes, in which simplistic assumptions are often made on the
physics in the core. Indeed, mixing is commonly considered to be
instantaneous and the most basic nuclear networks assume beryllium at its
equilibrium abundance. Those assumptions lead to significant differences in the
central composition of the elements for which the timescale to reach nuclear
equilibrium is lower than the convective timescale. In this work, we show that
those discrepancies impact the nuclear energy production and, therefore, the size
of convective cores in models computed with overshoot. We find that cores
computed with instantaneous mixing are up to 30\% bigger than those computed
with diffusive mixing. Similar differences are found when using basic nuclear
networks. Additionally, we observed an extension of the duration of the main
sequence due to those core size differences. We then investigated the impact of
those structural differences on the seismic modeling of solar-like oscillators.
Modeling two stars observed by \emph{Kepler}, we find that the overshoot
parameter of the best models computed with a basic nuclear network is
significantly lower, compared to models computed with a full nuclear network.
This work is a necessary step in improving the modeling of convective cores, which is
key to determining accurate ages in the framework of future space missions such as
Plato.}


\maketitle

\section{Introduction}
The extent of convective cores is still one of the most important open questions
in stellar physics. Classically, it is defined in stellar evolution codes
following the Schwarzschild criterion, which states that a region is convective
if $\gradrad > \gradad$, with $\gradrad$ and $\gradad$ being the usual radiative and
adiabatic gradients, respectively. This local and one-dimensional criterion
neglects several processes, such as overshooting, convective entrainment,
semi-convection, or rotational mixing -- which all contribute to extend the core
beyond the Schwarzschild boundary (i.e., beyond the shell where $\gradad =
\gradrad$).

Therefore, it is common practice to artificially increase the size of the core
over a certain distance, taken as a fraction of the pressure scale height. The
core boundary mixing can also be modeled as a diffusive process
\citep{freytag96} or as a turbulent entrainment \citep{Staritsin2013,Scott2021}.
In all cases, at least one free parameter is needed, known as the so-called overshoot
parameter. The tuning of this parameter can be done observationally, thanks to
the color-magnitude diagrams of open clusters
\citep[e.g.,][]{maeder89,vandenberg06}, modeling of binary stars
\citep[e.g.,][]{claret16}, or seismic modeling of main sequence stars
\citep[e.g.,][]{silva_aguirre13,deheuvels16,Mombarg2019,moravveji15,Pedersen2021}
and post-main sequence stars \citep{deheuvels11,Noll2021}. 

All those constraints rely on stellar evolution codes, which inevitably make
assumptions on the physics in the core. Regarding the nuclear reactions in the
core, it is usual to take some elements within the proton-proton (pp) chain,
such as beryllium, lithium, boron, and deuterium, at  chemical equilibrium.
Historically, this has been  done to make computations easier and it is justified by
the short nuclear timescales of those elements compared to the evolutionary
timescales \citep[e.g.,][]{Clayton1983}. Also, convective mixing is commonly
assumed to be instantaneous, meaning that all elements are homogeneous in a
convective region. It is for example the case in CESTAM \citep{Marques2013}, for
models without microscopic diffusion. This assumption is made to simplify the
computations, and is justified by the fact that the convective timescale is
small compared to the evolution timescale of the star. In this article, we
investigate the impacts of those two assumptions on the size of convective cores
in low-mass stars. Since are we focusing on stars with convective core, in which the
pp-chains are responsible for a significant part of the energy production, we
restrict our study to stars with masses between approximately $1.2$ and
$1.8\,\smass$.

In Sect.~\ref{physical_processes}, we recall the characteristics and assumptions
that are commonly done on the nuclear reactions and mixing in convective cores.
Then, in Sect.~\ref{impact_core_section}, we study how those assumptions may
impact the size of convective cores. In Sect~\ref{impact_seismic_modeling}, we
measure the impact of those core sizes differences on the seismic modeling of
main sequence solar-like oscillators. Finally,
Sect~\ref{discussions_conclusions} is dedicated to our discussions and
conclusions. 

\section{Nuclear reactions and mixing in convective cores}
\label{physical_processes}

Nuclear reactions and central mixing are two of the main physical processes that
define the core structure. In the following, to better understand the
characteristics of those two processes and how they interact, we briefly recall
their properties and timescales.

\subsection{Nuclear reactions}
\subsubsection{Characteristics}
\label{nucl_reac_carac}
Stars in the considered mass range (between approximately $1.2$ and
$1.8\,\smass$) produce a significant amount of their nuclear energy through the
proton-proton (pp) chain (see e.g. \citealt{kippenhahn,Clayton1983}). The
pp-chain is composed of several branches, the so-called pp-1, pp-2, and pp-3
branches, which represent different possible reactions to produce a $\heq$
nucleus. The rest of the energy is produced through the CNO-cycle.  

\begin{figure}
        \centering
        \includegraphics[width=0.5\textwidth, trim=0cm 2cm 0cm 0cm]{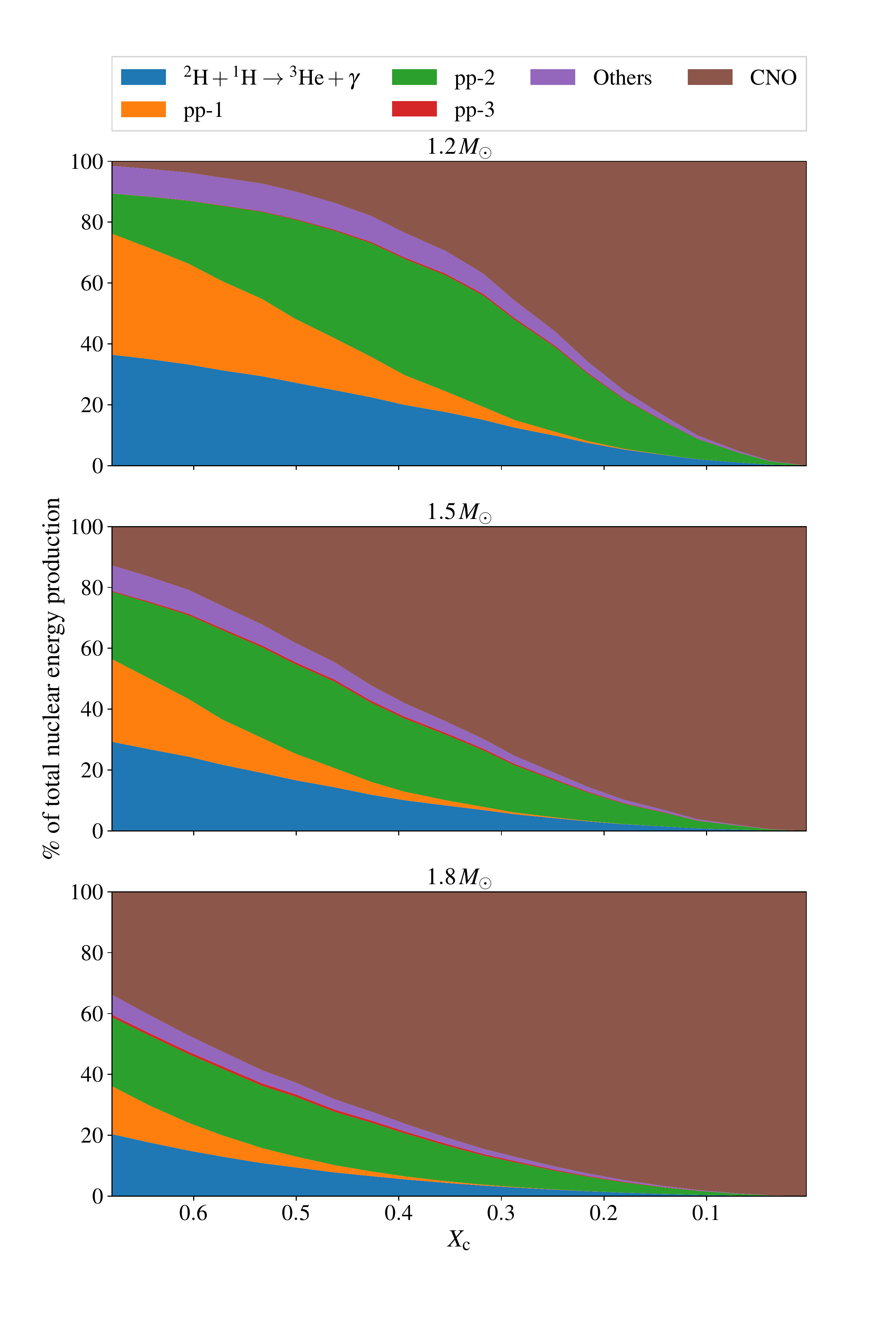}
        \caption{Share of the total energy production between the different
        reactions for models  of $1.2$, $1.5,$ and $1.8\,\smass$ stars at solar
        metallicity.}
        \label{energy_distrib}
\end{figure}

The part that the pp-chain takes in the total nuclear energy production, as well
as the branch that dominates, mainly depends on the composition and temperature
of the core. The mass and age of the star therefore have a significant impact.
This is illustrated in Fig.~\ref{energy_distrib}, which represents the evolution
of these distributions over the course of the main sequence, for $1.2$, $1.5,$
and $1.8\,\smass$ stars at solar metallicity, computed using the MESA stellar
evolution code \citep{Paxton2011,Paxton2013,paxton15}. We can see the gradually
increasing part of the CNO-cycle, with both the evolution and the mass, mainly
due to the increasing central temperature and the diminishing central hydrogen
abundance. 

Moreover, we may note that the part produced by the pp-chain is (mostly) divided
between three reactions: 1) $\deut + \h \rightarrow \het + \gamma$, which rather
dominates at low temperature; 2) The pp-1 branch,  $\het + \het \rightarrow \heq
+ 2\,\h$, which is the dominant way to produce $\heq$ at low temperature; 3) The
pp-2 branch, whose energy production is dominated by the contribution of the
$\li + \h \rightarrow 2 \,\heq$ reaction.

\subsubsection{Timescales}
To better determine the characteristics of those reactions and compare their
interactions with other physical processes, we computed in the following their
timescales. We defined the nuclear chemical timescale, $\tau_i^{\mathrm{nucl}}$,
of an element $i$, similarly to \cite{Clayton1983}, by:
\begin{equation}
        \left| \left( \frac{\partial X_i }{\partial t} \right)_{\mathrm{nucl}} \right| = \frac{X_i}{\tau_i^{\mathrm{nucl}}}, 
\end{equation}
where $X_i$ is the mass fraction of the element. Combining this definition with
the expression of $ (\partial X_i / \partial t)_{\mathrm{nucl}}$ (e.g., Eq.~8.4
from \citealt{kippenhahn}), we find:

\begin{equation}
        \tau_i^{\mathrm{nucl}} = \frac{X_i \rho}{m_i} \left| \sum_j r_{j} - \sum_k r_k \right|^{-1},
\end{equation}
where $\rho$ is the density, $m_i$ is the atomic mass of the element, $r_{j}$ is
the reaction rate (number of reactions per unit of volume and time) of the
reactions in which the element $i$ is a product and $r_k$ represents the
reactions in which it is a reactant. Using MESA, we computed the values of
$\tau^{\rm nucl}$ averaged over the core, for different elements, of a $1.3\,
\smass$ star at solar metallicity evolved up to the moment when the central
hydrogen mass fraction is $X_c=0.5$. Our results are listed in
Table~\ref{table_timescale}.

    \begin{table}
                \caption{Nuclear timescales of different elements for a $1.3\,\smass$ star at solar metallicity, evolved up to $X_c=0.5$ }
        \centering
        \begin{tabular}{lc}
                \hline \hline
                Element & Nuclear Timescale \\
                \hline
                $\h$ & 2.31 Gyr \\
                $\deut$ & 0.891 s \\
                $\het$ & $ 2.11 \times 10^5$ yr \\
                $\heq$ & 17.6 Gyr \\
                $\li$ & 7.48 hours \\
                $\be$ & 1.60 yr \\
                $\bh$ & 156 s \\
                \hline          
        \end{tabular}
        \label{table_timescale}
    \end{table}

\subsection{Central mixing}
\subsubsection{Characteristics}
Within the considered mass range, central mixing is attributed to convection. In
stellar evolution codes, it is generally modeled in an instantaneous or
diffusive way. In the first case, all elements are considered to be homogeneous
in the core at every time step. This is, for example, the case of CL\'ES
\citep{Scuflaire2008}, ASTEC \citep{jcd2008-astec}, and CESAM2k/CESTAM
\citep{morel08,Marques2013} models without microscopic diffusion. In the second
case, the mixing is modeled as a very efficient diffusion process. Such
implementation is found in MESA, STAROX \citep{Roxburgh2008}, optionally in
GARSTEC \citep{Weiss2008}, and in CESAM2k or CESTAM models with microscopic
diffusion. In the three first cases, the diffusion coefficient is computed with
the mixing-length theory. For CESAM/CESTAM with microscopic diffusion, it is
fixed at $10^{13}\,\mathrm{cm}^2.\mathrm{s}^{-1}$, which allows for  very
efficient mixing and a good numerical stability.

In this work, we use the CESTAM code without microscopic diffusion to
model stars with an instantaneous mixing, along with the MESA code to model stars with
a diffusive mixing. We ensured that the two codes use similar physics, and
(especially) the same nuclear cross-sections. The rates are taken for both code
from the NACRE compilation \citep{Angulo1999} except for the \element[][14]{N} +
\element[][1]{H} reaction, for which the rate is taken from \cite{Imbriani2005}.
Moreover, the two codes use the same step overshooting distance, $\dov$,
prescription, namely, $\dov = \ov \min(H_p, R_\mathrm{cc})$, with $\ov$  as a free
parameter, $H_p$ as the pressure scale height, and $R_\mathrm{cc}$ as the radius of the
convective core. Finally, both codes use the same solar abundances (coming from
\cite{asplund09}), opacity tables (namely OPAL, \citealt{opal_opacities}), and
convection model, namely: the mixing length theory. 

\subsubsection{Timescales}

In the case of an instantaneous mixing, the timescale is constant and equal to
zero. In the case of a diffusive mixing, the timescale is given by:
\begin{equation}
        \tau_{\mathrm{conv}} = \int_0^{R_{\mathrm{cc}}} \frac{\dd r}{v_{\mathrm{conv}}},
\end{equation}
with $R_{\rm cc}$ as the radius of the convective core and $v_{\mathrm{conv}}$ as the
convective velocity determined from the mixing-length theory. Using the same
model as in Sect.~\ref{nucl_reac_carac}, we find $\tau_{\rm conv} =
43\,\mathrm{days}$.

\subsection{Impact of instantaneous mixing on central composition}
When comparing the values from Table~\ref{table_timescale} with $\tau_{\rm
conv}$, we can note that the nuclear timescale is lower than $\tau_{\rm conv}$
for $\deut$, $\li,$ and $\bh$. This means that they reach their mass fraction
equilibrium value faster than they are mixed by convection; consequently,
they are not homogeneous in the core. On the contrary, $\h$, $\het$, $\heq$, and
$\be$ have a higher nuclear timescale than the convective timescale and,
consequently, they are efficiently mixed and then homogeneous in the core. 

\begin{figure}
        \centering
        \includegraphics[width=0.5\textwidth]{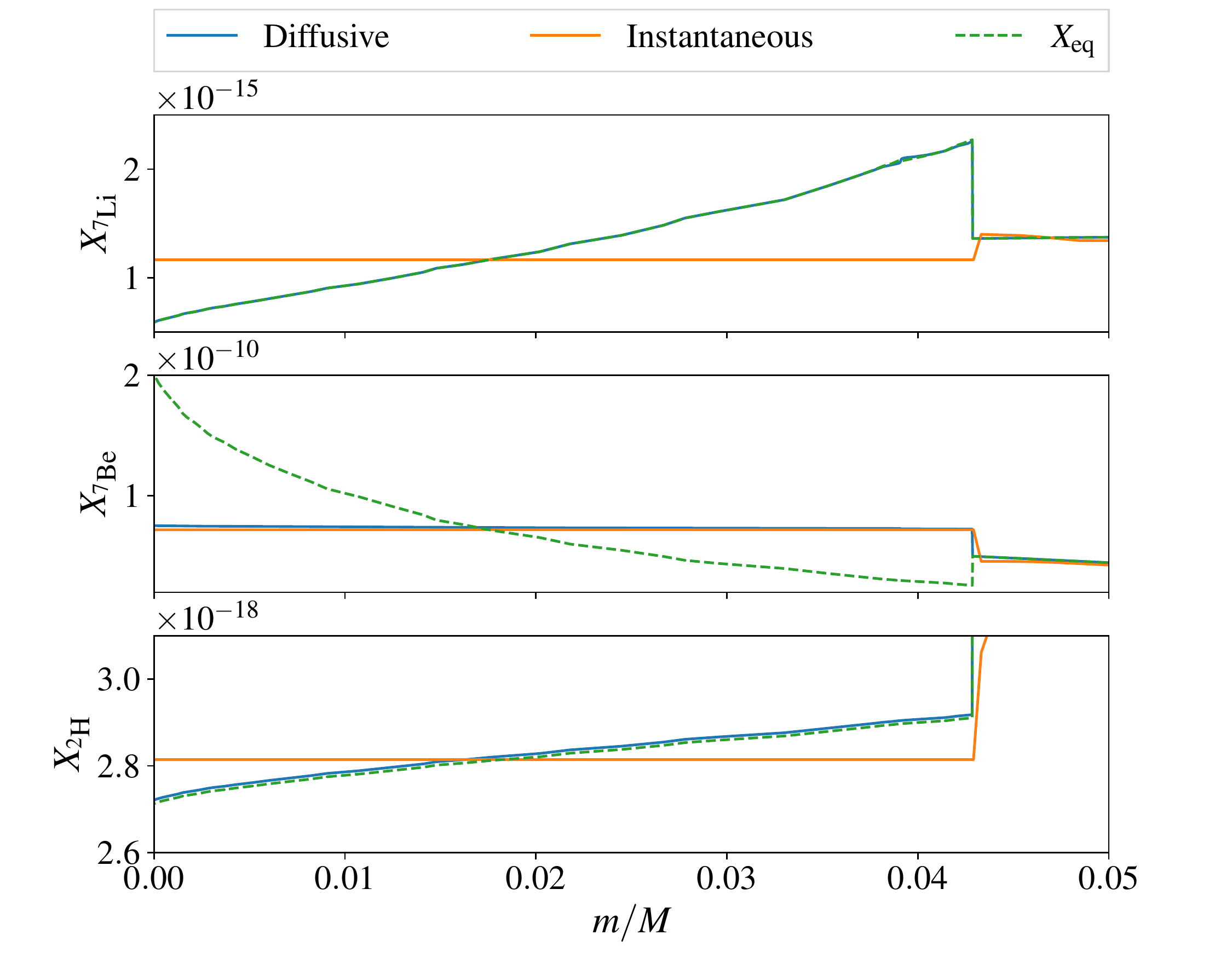}
        \caption{Abundance profiles in the core of a $1.3\,\smass$ stellar model,
        evolved until $X_c = 0.5$ and computed either with MESA (central diffusive
        mixing) or CESTAM (central instantaneous mixing). We also represent the
        equilibrium abundance profiles in dotted green line.}
        \label{compare_elements}
\end{figure}

We represent in Fig.~\ref{compare_elements} the mass fractions of $\li$, $\be,$
and $\deut$ in the core of main sequence models computed either with MESA (using
a diffusive mixing) or CESTAM (using an instantaneous mixing). We also represent
the composition profiles of the elements that are obtained when their nuclear
equilibrium is assumed, that is, $X_{i,\mathrm{eq}}$ such that $(\partial X_i /
\partial t)_{\mathrm{nucl}} = 0$. In models with an instantaneous mixing, all
the elements are by construction homogeneous in the core. This is in conflict
with what we could expect from the timescale comparison: indeed, both lithium
and deuterium should reach their equilibrium abundances. Therefore, we conclude
that those compositions are incorrect. In models with diffusive mixing, however,
we find the expected behavior: both lithium and deuterium are at the equilibrium
abundance and beryllium is almost homogeneous. 

\subsection{Case when  $\deut$, $\be$, $\li,$ and $\bh$ are assumed to be at equilibrium}
\label{assuming_simple_networks}

In stellar evolution codes, it is often assumed that $\deut$, $\be$, $\li,$ and
$\bh$ are at equilibrium at all times. This is for instance the default choice
in MESA, through the parameter \verb|default_net_name = 'basic.net'|. In the
following we refer to those simplistic nuclear networks as ``basic'' networks,
in opposition to the ``full'' networks that take into account all the nuclear
timescales. Using a basic network is equivalent to assuming $\tau^\mathrm{nucl}
= 0$ for the affected elements and is justified by their short nuclear
timescale compared to the typical nuclear timescale of the star (see
Table~\ref{table_timescale}). However, for beryllium, we note that
$\tau^\mathrm{nucl}_{\be} > \tau_\mathrm{conv}$ and, thus, that this element is
efficiently mixed. Therefore, using a basic network yields a wrong beryllium
composition profile, as we can see in Fig.~\ref{compare_elements} in which the
equilibrium composition profile of this element differs from the efficiently
mixed one. This faulty beryllium composition impacts in turn the lithium
composition, since the latter is produced by the electronic capture of the former. Consequently, we can expect effects on the core structure that are
comparable to those caused by an instantaneous mixing.

\section{Impact on core structure}
\label{impact_core_section}
We observed in the previous section that using an instantaneous mixing, or a
basic nuclear reaction network, has an impact on the central composition of the
star. In this Section, we investigate how this change of composition impacts the
structure of the convective core.

\subsection{Nuclear energy production}

\begin{figure*}
        \centering
        \includegraphics[width=\textwidth]{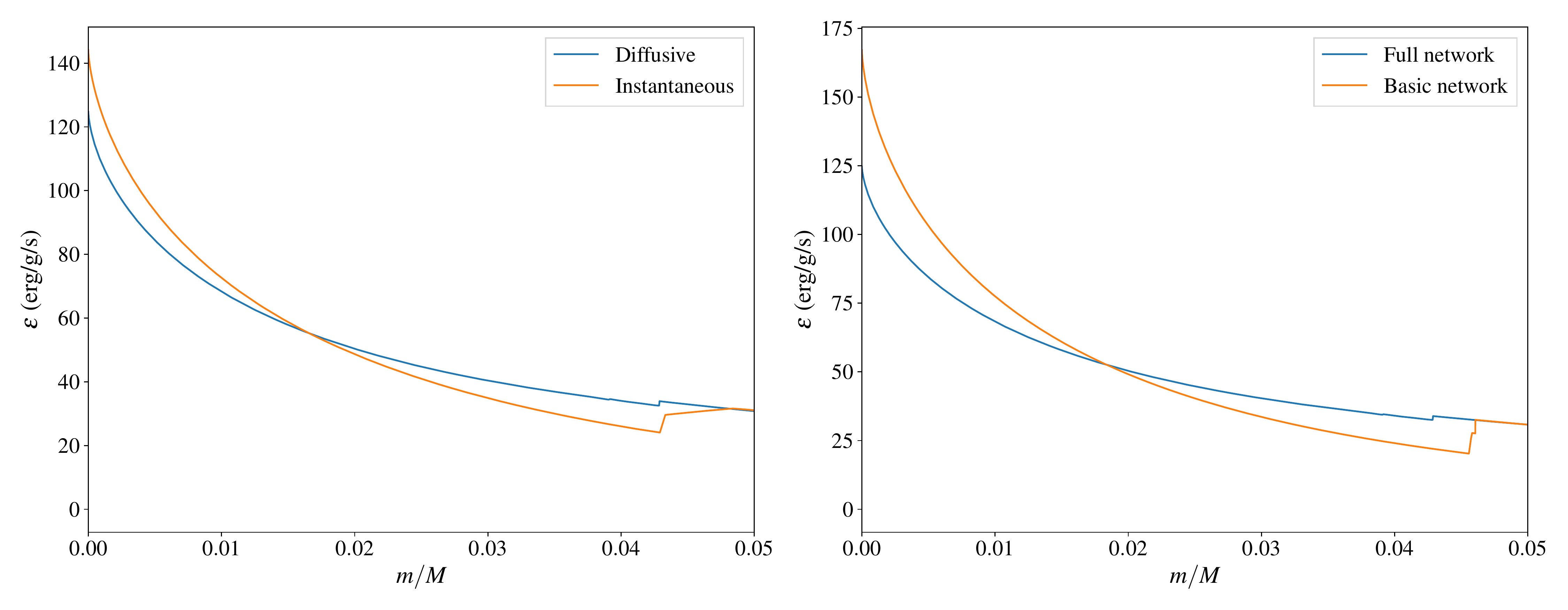}
        \caption{Rate of energy generation per unit mass. \emph{Left Panel}: Models
computed with either diffusive (blue) or instantaneous (orange) mixing.
        \emph{Right Panel}: Models computed with either the full (blue) or basic
        (orange) network.}
        \label{epsilon_profile}
\end{figure*}
The nuclear energy generation rate per unit mass (noted as $\epsilon_{ij}$ for a
reaction involving two reactants $i$ and $j$) is related to their mass fractions
through the expression:

\begin{equation}
        \label{eq_eps}
    \epsilon_{ij} = \frac{1}{1 + \delta_{ij}} \frac{Q_{ij}}{m_i m_j}\rho X_i X_j \langle \sigma v \rangle_{ij},
\end{equation}
where $\delta_{ij} \equiv 1$ if the two reactants are identical, 0 otherwise,
$Q$ the energy released per reaction, $\sigma$ is the cross-section of the
reaction, and $v$ is the velocity of the particles involved in the reaction. We
can conclude from this expression that the observed differences in
Fig.~\ref{compare_elements} should have an impact on the nuclear energy
production of the star. Moreover, this effect should not be negligible, as the
reaction involving $\li$ as a reactant can represent around 20\% of the total
nuclear energy production for stars with masses between $1.2$ and $1.8\,\smass$
(see Fig.~\ref{energy_distrib}). This is indeed verified in the models. Thus,
models computed with different mixing prescriptions yield different $\epsilon$
profiles, as shown in the left panel of Fig.~\ref{epsilon_profile}. Using a
different nuclear network impacts $\epsilon$ comparably, as shown in the right
panel of Fig.~\ref{epsilon_profile}. We note, however, that when integrating
$\epsilon$ over the core mass, we find similar core luminosities for the two
models. 

Additionally, we may wonder about the impact of the differences in deuterium
abundance on the total energy production. Indeed, the deuterium composition
profile differs between models with different mixing (see
Fig.~\ref{compare_elements}). Moreover, the deuterium is a reactant of a
reaction ($\deut + \h \rightarrow \het + \gamma$), which represents a significant
part of the total energy production (see Fig.~\ref{energy_distrib}). Yet,
contrary to the case of  lithium, the relative differences in abundance between models
with diffusive and instantaneous mixings are small (around 3\%). This is due to
the fact that the reactions involving deuterium are less sensitive to
temperature than the ones involving lithium. Therefore, the equilibrium
abundance of deuterium does not, in a relative sense, radially vary as much as
the lithium abundance in the core and is less impacted by the instantaneous
mixing. The impact of this difference on the total nuclear energy production is,
thus, negligible.

\subsection{Convective core size}
For main sequence stars, which are at thermal equilibrium, the integral of
$\epsilon$ over the mass inside a shell located at radius $r$ is the luminosity
going through that shell, noted as $L_r$. We know that the radiative gradient,
$\gradrad = (\partial \ln T / \partial \ln p)_{\rm rad}$, is related to $L_r$
through the expression:
\begin{equation}
        \gradrad = \frac{3}{16\pi a c G}\frac{\kappa L_r p}{m T^4},
\end{equation}
where $a$ is the radiation density constant, $c$ is the velocity of light, $G$ is the
gravitational constant, $\kappa$ is the Rosseland opacity, $p$ is the pressure, $T$
is the temperature, and $m$ is the mass within the sphere at the radius, $r$.

\begin{figure*}
        \centering
        \includegraphics[width=\textwidth]{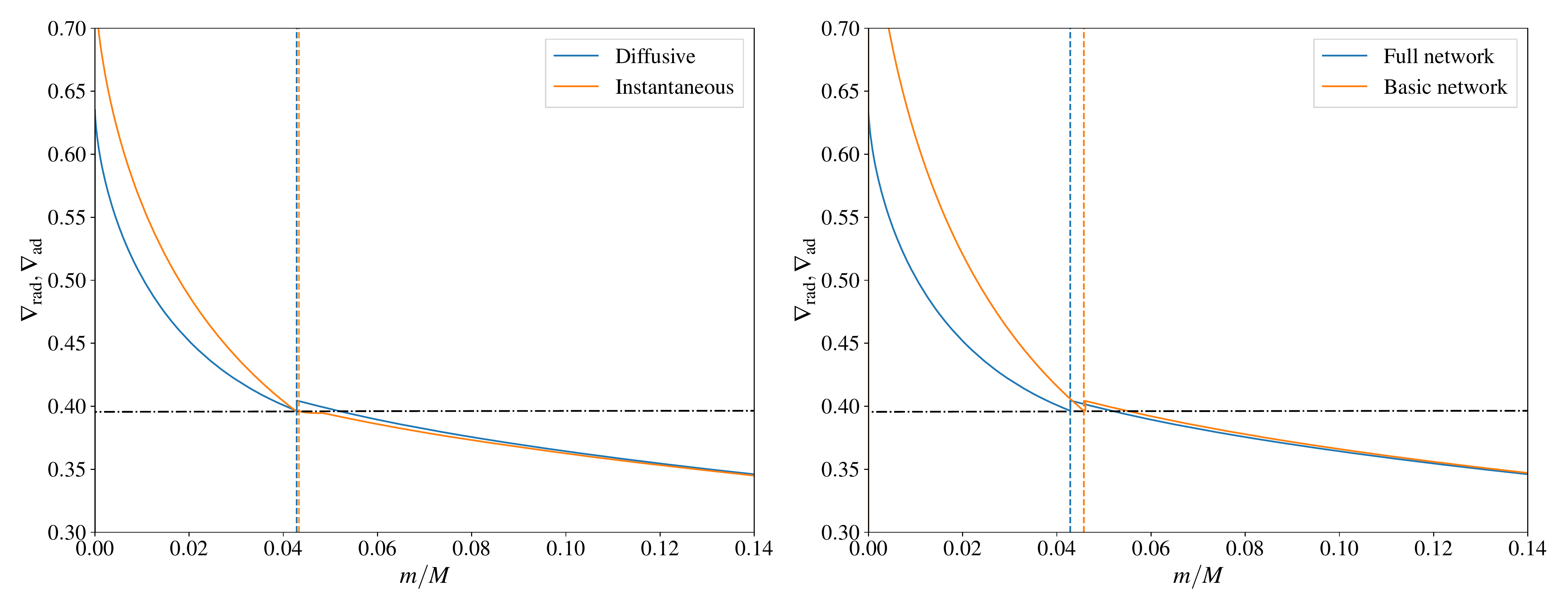}
        \caption{Radiative gradients of $1.3\,\smass$ of models, stopped at
        $X_c=0.4$. The adiabatic gradient is represented in dot-dashed black line.
        The vertical lines represent the core boundaries. \emph{Left Panel}: Models
computed with either diffusive (blue) or instantaneous (orange) mixing.
        \emph{Right Panel}: Models computed with either the full (blue) or basic
        (orange) network.}
        \label{rad_grad_prof_no_ov}
\end{figure*}

\begin{figure*}
        \centering
        \includegraphics[width=\textwidth]{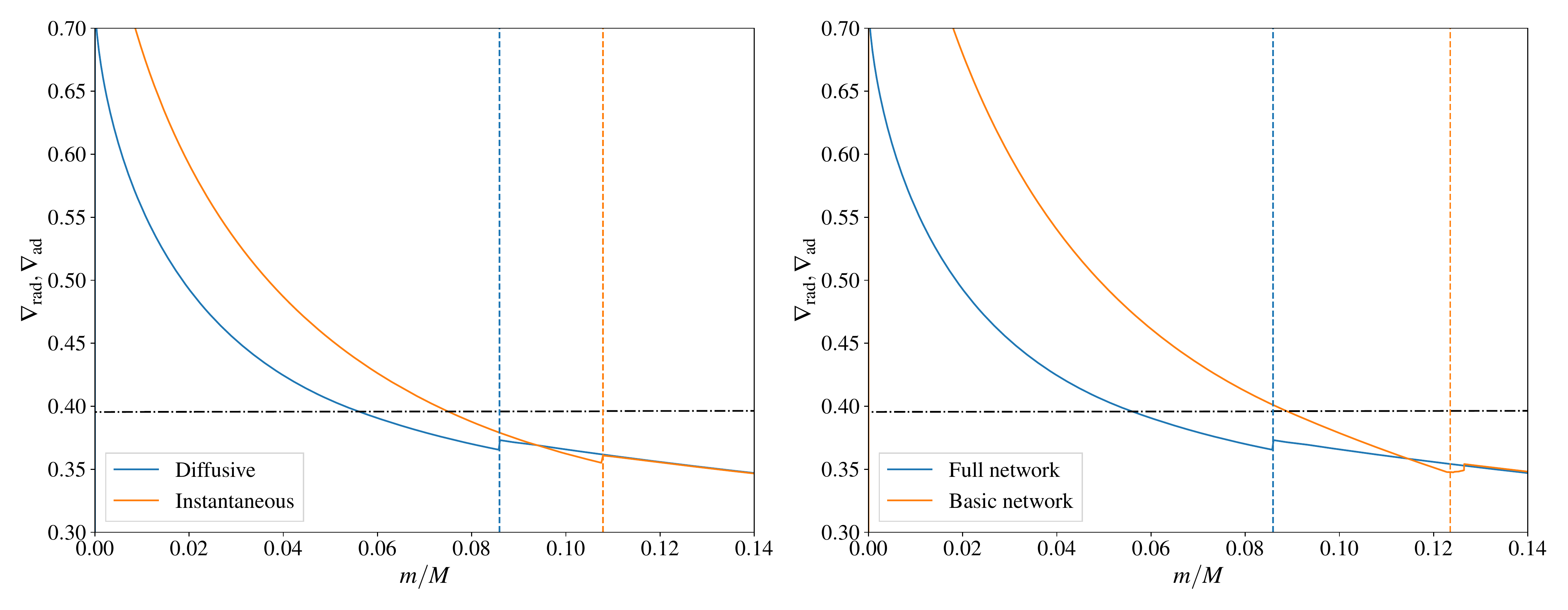}
        \caption{Same as Fig.~\ref{rad_grad_prof_no_ov}, but the models are computed
        with $\ov=0.15$.}
        \label{rad_grad_prof_ov}
\end{figure*}

\begin{figure}
        \centering
        \includegraphics[width=0.5\textwidth]{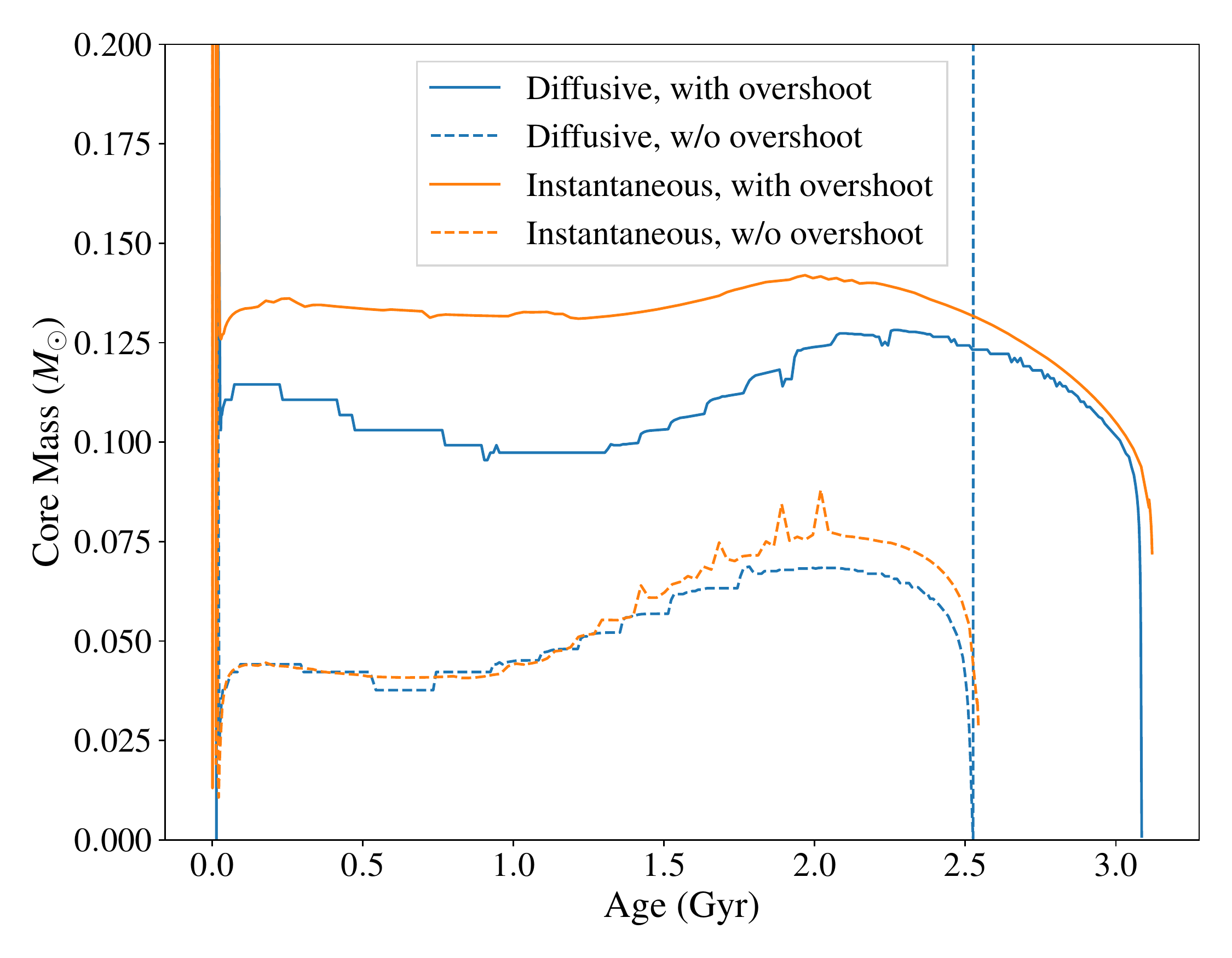}
        \caption{Core mass evolution of $1.3\,\smass$ models computed either with a diffusive (blue) or an instantaneous (orange) mixing, with (full, $\ov = 0.15$) or without (dotted) overshooting.}
        \label{evol_core_mass}
\end{figure}

\begin{figure}
        \centering
        \includegraphics[width=0.5\textwidth]{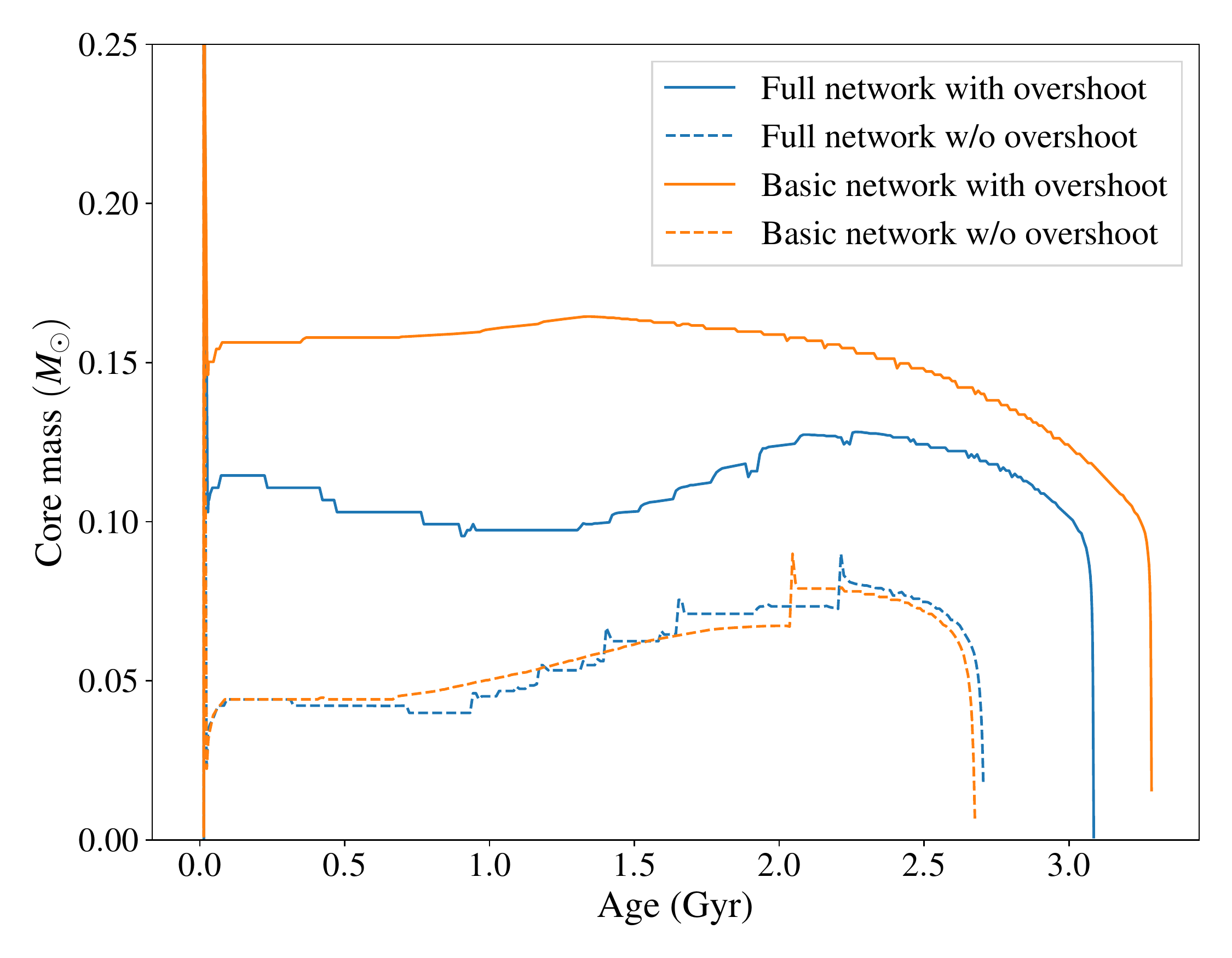}
        \caption{Core mass evolution of $1.3\,\smass$ models, using either a full (blue) or basic (orange) nuclear reaction network and computer with (full, $\ov = 0.15$) or without (dotted) overshoot.}
        \label{evol_core_basic_net}
\end{figure}

Therefore, the difference in $X_{\li}$ has an impact, through the $L_r$ term, on
the radiative gradient in the core. Figs.~\ref{rad_grad_prof_no_ov} and
\ref{rad_grad_prof_ov} illustrate this aspect, respectively, without and with overshoot
($\ov = 0.15$). The different profiles of $\gradrad$ have another striking
effect on the stellar structure: for the models with overshoot, the model with
an instantaneous mixing (or basic network) exhibits a bigger convective core
than the model with a diffusive mixing (or full network). Such behavior is
observed during the whole main sequence, as we can see in
Fig.~\ref{evol_core_mass}: the models with instantaneous mixing having a core up
to 30\,\% more massive than the cores computed with a diffusive mixing. A similar
behavior is shown in Fig.~\ref{evol_core_basic_net} for models using a basic
network, which exhibit more massive cores than models using a full network.

We may note that a small region situated just outside the convective core in
the MESA model is semi-convective, in the sense that it is unstable according to
the Schwarzschild criterion, but stable according to the Ledoux criterion. This
region has small to no impact on the stellar evolution and is quickly erased if
a tiny amount of overshoot is added. Moreover, it does not exist in the CESTAM
model due to the numerical scheme that smooths out the composition profile and
thus lowers the opacity jump. Consequently, this semi-convective region does not
impact the conclusions of this paper. 


\subsection{Link with core overshooting}
\label{why_only_overshoot}
\begin{figure}
        \centering
        \includegraphics[width=0.5\textwidth]{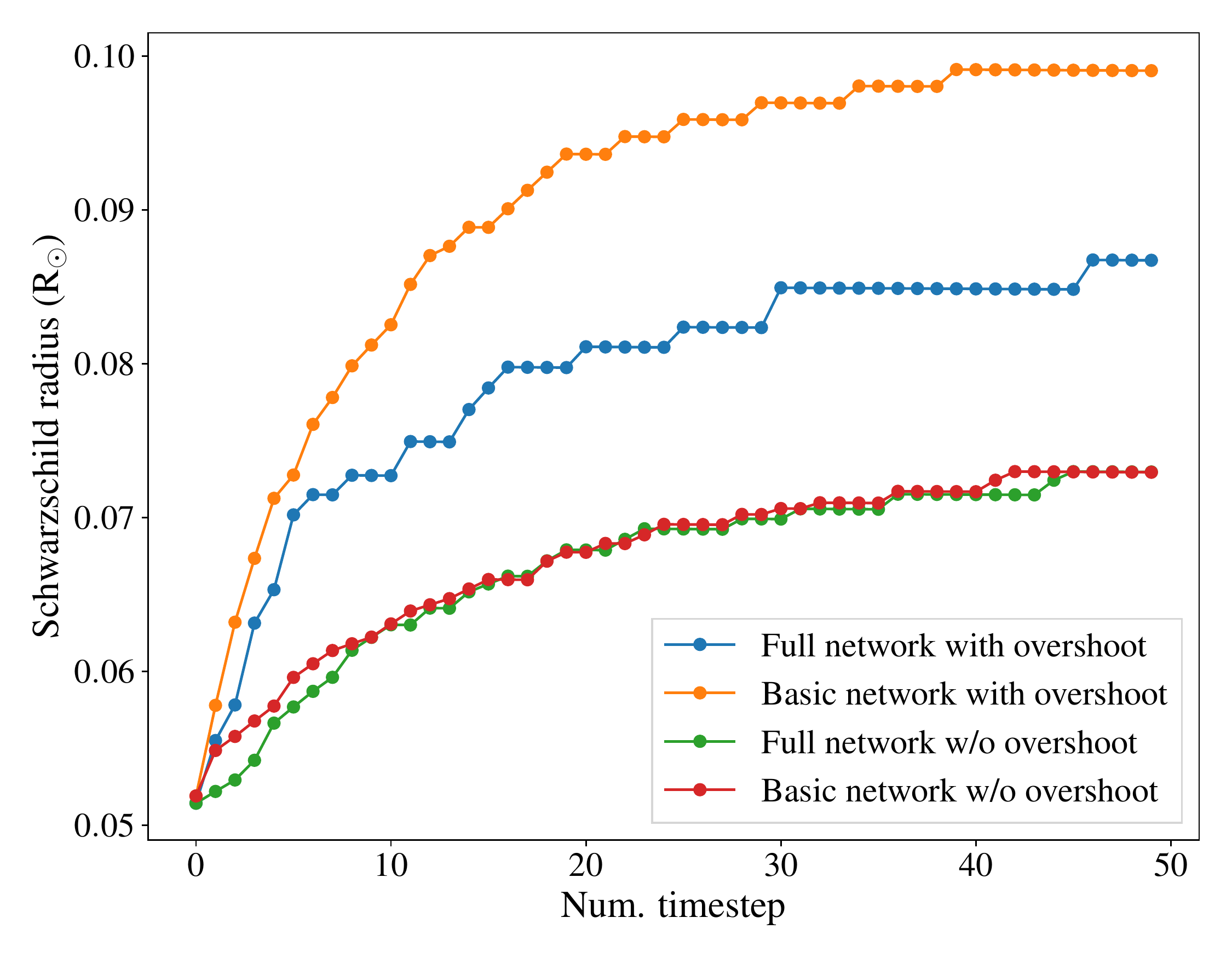}
        \caption{Evolution of the Schwarzschild radius of the convective core at the very beginning of the main-sequence, for models with a full nuclear reaction network and overshoot (blue), a basic reaction network and overshoot (orange), a full network without overshoot (green), and a basic network without overshoot (red). The 50 first time steps correspond to approximately 50 million years.}
        \label{evol_core_timestep}
\end{figure}
To better understand why only models with an extension of the convective core
have a significantly different core size, we investigated the evolution of the
convective core at the very beginning of the main sequence. To do so, we added
overshooting to the core just after the initial contraction phase, once the star
reached a thermal equilibrium state. Figure~\ref{evol_core_timestep} shows the
evolution of the Schwarzschild radius (i.e., the radius where $\gradrad =
\gradad$) for models computed with and without overshoot, and with different
nuclear reactions networks, for the first 50 timesteps of the main-sequence.
Similarly to models presented before, changing the nuclear reactions networks
only impacts the Schwarzschild radius of models with overshoot. We noted that
this process happens in a given number of time steps, rather than a given time
(hence the choice of the x-axis in Fig.~\ref{evol_core_timestep}), which
indicates that it is numerical rather than evolutionary. To better understand
it, we introduce a toy model that mimics the behavior of the core boundary
during the first time steps that follow the addition of overshooting.

\begin{figure}
        \centering
        \includegraphics[width=0.45\textwidth]{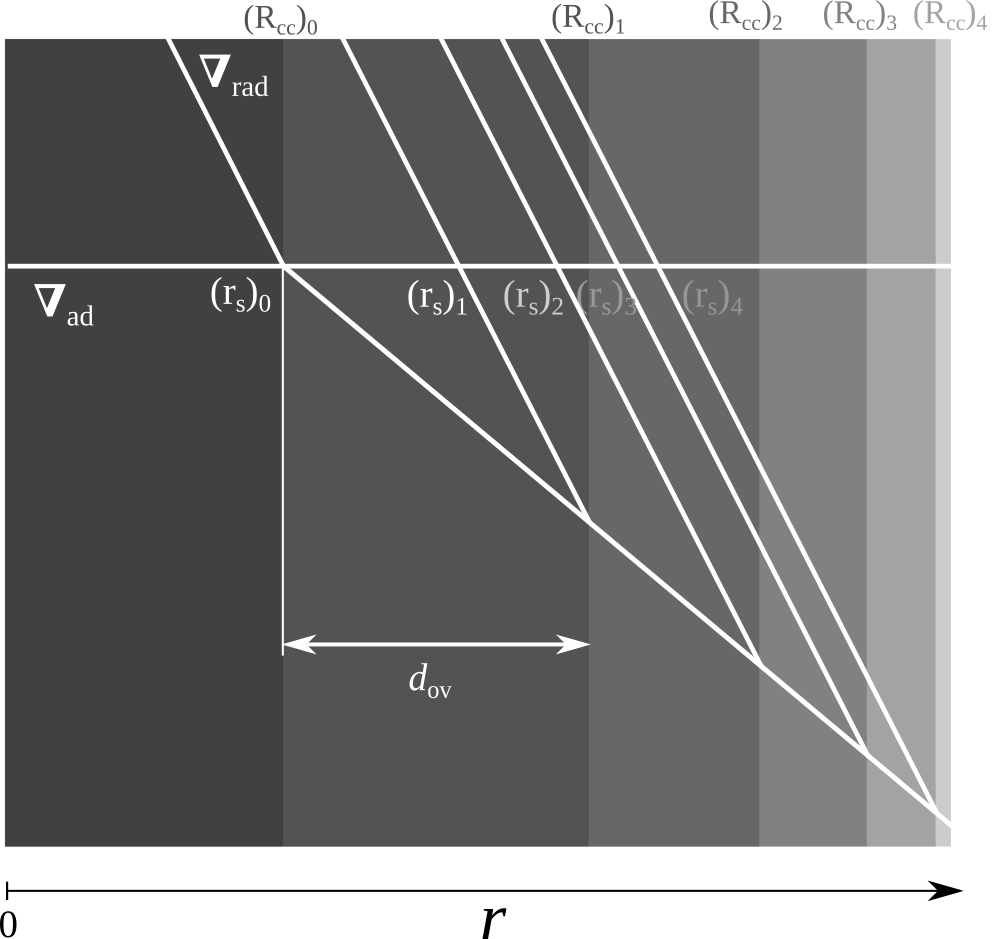}
        \caption{Schematic representation of the toy-model explained in
        Sect.~\ref{why_only_overshoot}.}
        \label{cartoon}
\end{figure}

In this toy model, which is illustrated in Fig.~\ref{cartoon}, we simplified the
radiative gradient near the boundary of the convective core as a piecewise
linear function of slopes $a'$ in the core and $a$ in the radiative region. The
adiabatic gradient is assumed to be constant. The Schwarzschild radius, where
$\gradrad = \gradad$, is noted as $r_s$ and the boundary of the fully mixed region
is noted as $\rcc$. The time step number is noted with a subscript.

At first, the core is not extended, therefore $(r_s)_0 = (\rcc)_0$. We then
add overshoot over a distance $\dov$ such that $(\rcc)_1 = (r_s)_0 + \dov$. At
the next timestep, the Schwarzschild radius is defined by the intersection of
$\gradrad$ with $\gradad$ and, therefore, $(r_s)_1 > (r_s)_0$, due to the fact
that the slope of $\gradrad$ is different in the core and in the radiative
region. Then, overshoot is added again (we take as an approximation a constant
$\dov$) such that $(\rcc)_2 = (r_s)_1 + \dov$. The process is repeated at the next
timestep, and $(r_s)_3 > (r_s)_2 > (r_s)_1$. Eventually, this converges to a
Schwarzschild radius that is significantly larger, as observed in
Fig.~\ref{evol_core_timestep}. 

This simple toy model allows us to derive an analytical expression for the final
Schwarzschild radius, once the process is converged, that we note
$(r_s)_\mathrm{f}$. We find:
\begin{equation}
        (r_s)_\mathrm{f} = (r_s)_0 +  \dov \left(\frac{a'}{a} - 1 \right).
\end{equation}

We can notice that $(r_s)_\mathrm{f}$ depends on the slopes of $\gradrad$ in
the core and in the radiative region. Yet, the slope of $\gradrad$ in the core
differs between models with different mixing or nuclear reaction networks (see
Fig.~\ref{rad_grad_prof_no_ov}), which causes $(r_s)_\mathrm{f}$ of those models
to differ as well. If we note $\alpha$ and $\beta$ as the slopes of $\gradrad$ in
the core for two different models, assuming that they have the same radiative
gradient in the radiative region (as seen in models, e.g. in
Fig.~\ref{rad_grad_prof_no_ov}), we find that the difference of
$(r_s)_\mathrm{f}$,  noted as $\Delta r_s$, is equal to:
\begin{equation}
        \label{toy_model_equation}
        \Delta r_s = \dov \left( \frac{\alpha - \beta}{a} \right).
\end{equation}
This explains the different convective core sizes for models with overshoot.
Conversely, if the model has no overshoot, the different radiative gradient
slopes do not impact the size of the core.

\subsection{Impact on the duration of the main sequence}

\begin{figure}
        \centering
        \includegraphics[width=0.5\textwidth]{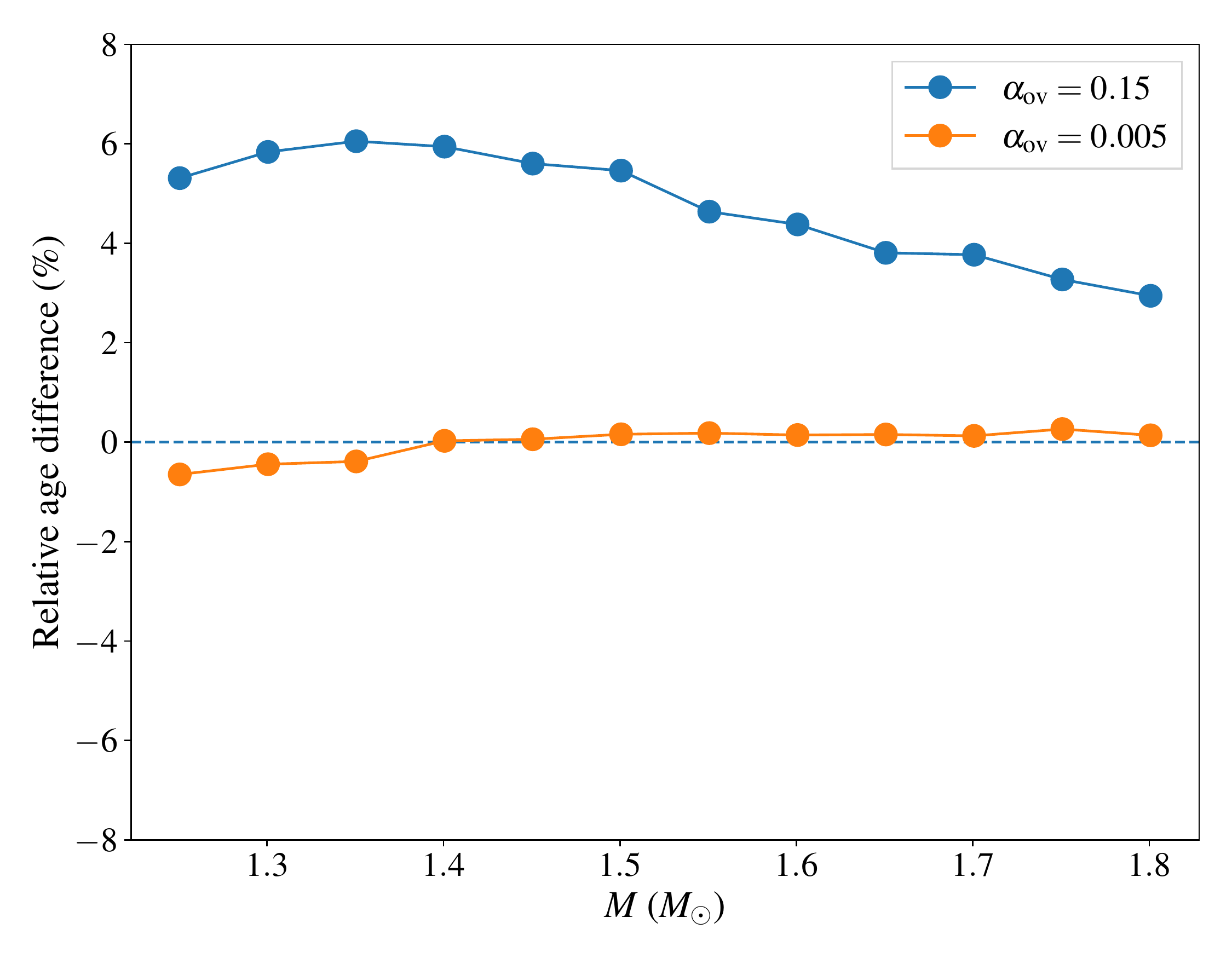}
        \caption{Relative age difference (in the sense basic - full) of models evolved until $X_c = 0.001$, computed with $\ov = 0.15$ (blue) and $\ov = 0.005$ (orange).}
        \label{age_ms_diff}
\end{figure}

It is well known that the size of the convective core impacts the duration of
the main sequence. As the process described in this paper increases the size of
the convective core, we investigated the change in the duration of the main
sequence when using a basic nuclear reaction network rather than a full network.
To do so, we computed models with different networks that evolved until the end
of the main-sequence (arbitrarily defined as $X_c=0.001$), varying their masses.
We then compared their age. Fig.~\ref{age_ms_diff} represents the relative age
differences, in the sense of basic minus full. We observe that models computed using a
basic network have a main sequence that is between 3-6\% longer, depending
on the mass. In order to ensure that those differences are due to the difference
of core size and not the energy production, we represent the same differences
computed for models with almost no overshoot\footnote{Models with $\ov = 0$ have
a more erratic core size evolution (see Fig.~\ref{evol_core_mass}), which
impacts the main sequence duration, making the comparison less clear (see
\citealt{Lebreton2008} for examples of erratic core boundary behavior for
several evolution codes.)},  $\ov = 0.005$. Indeed, those models are supposed to
have the same core size (see Fig.~\ref{evol_core_mass}). For those models, no
age difference is found, which confirms that those found for models with $\ov =
0.15$ result from the process described in this paper.

This difference in main sequence duration may impact the modeling of post-main
sequence stars, inducing systematic biases in the age determination. We note
that the observed differences are comparable with uncertainties inferred through
seismic modeling of subgiant stars, for instance (e.g., \citealt{Noll2021}).

\section{Impact on the seismic properties of solar-like oscillators}
\label{impact_seismic_modeling}
Stars with masses below approximately $1.5\,\smass$ are solar-like oscillators.
Therefore, they exhibit numerous pressure (p) modes, which allow us to probe the
internal structure of the star. Those modes are mainly sensitive to the upper
structure of the star (see e.g., \citealt{Aerts2010}), but they are also able to
probe the central region of the star. We could therefore expect that the
differences in the convective core sizes observed in this work may impact the
seismic observables. Thus, in this section, we describe our seismic study of
solar-like pulsations using either a full nuclear reaction or a basic nuclear
reaction network. We compare nuclear reaction networks rather than mixing
prescriptions, because the latter would require comparisons between two
different stellar evolution codes, which could potentially induce other biases. 

In this section, we first study the impact of the nuclear network on the seismic
observables of a grid of models. Then, we model two stars observed by
\emph{Kepler} to quantify the impact of using a basic reaction network on the
retrieved stellar parameters.

\subsection{Method}
\subsubsection{Characteristics of the models}
\label{carac_model}

All the models were computed using the MESA v10108
\citep{Paxton2011,paxton15,paxton18} stellar evolution code. We used the OPAL
equation of states and opacity tables \citep{opal_eos,opal_opacities}, with a
solar mixture from \cite{asplund09}. Convective regions were computed following
the mixing-length theory prescription of \cite{cox68}. The mixing-length
parameter $\conv$ has been fixed to a solar-calibrated value of 1.9. Microscopic
diffusion has not been taken into account. Overshooting was modeled as step
extension of the convective core, with chemical mixing only: the temperature
gradient is equal to the radiative gradient outside the Schwarzschild limit. The
distance over which the core is extended is taken as:
\begin{equation}
        \label{def_ov_mesa}
        d_{\mathrm{ov}} = \ov \min \left( H_p, R_{\mathrm{cc}} / \conv \right).
\end{equation}

This definition corresponds to the default definition in MESA, and differs from
the one used in Sect.~\ref{impact_core_section}, which is the default in
CESTAM. Also, we note that in the considered range of parameters, convective
cores are small, which implies that $H_p > R_{\mathrm{cc}} / \conv$ at the
Schwarzschild radius for most of the models. The distance of overshooting is
therefore generally defined as $d_{\mathrm{ov}} = \ov R_{\mathrm{cc}} / \conv$
in the considered mass range.

Finally, the adiabatic oscillations have been computed using ADIPLS
\citep{jcd08}. No surface effect correction has been applied, as we use seismic
indicators which are almost independent of the upper layers of the star (see
Section~\ref{seismic_obs}). 

\subsubsection{Characteristics of the grid}
\label{carac_grid}
\begin{table}
        \caption{Parameter space of the grid.}
        \centering
        \begin{tabular}{c c c}
                \hline\hline
                Param. & Min & Max \\
                \hline
                $M$ ($\smass$) & 1.1 & 1.45 \\
                $[Z/X]$ (dex) & -0.2 & 0.35 \\
                $Y_0$ & 0.24 & 0.33 \\
                $\ov$ & 0 & 0.45 \\
                \hline          
        \end{tabular}
        \label{table:param_grid}
\end{table}

We computed two grids, one using a basic network and the other one using a full
reaction network. Using Sobol sequences, we computed  for each grid $2^{13} =
8192$ uniformly distributed tracks. The varying parameters are the stellar mass,
$M$, the metallicity, $[Z/X]$, the initial helium abundance, $Y_0$, and the
overshoot parameter. The parameter space is detailed in
Table~\ref{table:param_grid}. For every track, 60 models have been computed
during the main sequence, evenly spaced in central hydrogen abundance.

\subsubsection{Definition of the seismic observables}
\label{seismic_obs}

The process studied in this article only impacts the core structure and has
negligible effect on the global structure of the star. Therefore, we focused on
seismic observables that are sensitive to the core structure. The small
separations between the modes of degrees $l=0$ and 1 were shown to be the most
sensitive to the core size \citep{Provost2005}. Moreover, \cite{Roxburgh2003}
showed that the $r_{01}$ ratio is nearly insensitive to the surface layers of
the star and therefore to the so-called near-surface effects. It is defined as:
\begin{equation}
        r_{01}(n) = \frac{\nu_{n-1,0} - 4\nu_{n-1,1} + 6\nu_{n,0} - 4\nu_{n,1} + \nu_{n+1,0}}{8 \Delta \nu_1(n)},
\end{equation}
where $\nu_{n,l}$ is the frequency of the mode of radial order $n$ and degree
$l$ and $\Delta \nu_1 (n) = \nu_{n,1} - \nu_{n-1, 1}$. Finally, following
\cite{deheuvels16}, we fit to $r_{01}$ a polynomial of degree two of the type:
\begin{equation}
        P(\nu) = a_0 + a_1(\nu - \beta) + a_2(\nu - \gamma_1)(\nu - \gamma_2),
\end{equation} 
and use the coefficients $a_0$, $a_1$ and $a_2$ as observables. Especially,
$a_0$ and $a_1$ are known to be sensitive to the size of the mixed core
\citep{Silva-Aguirre2011,deheuvels16}. The addition of three more degrees of
freedom, namely the coefficients $\beta$, $\gamma_1$, and $\gamma_2$, allows to
ensure the independence of the $a_k$ coefficients. This permits the use of
classical $\chi^2$ minimization methods (see Section~\ref{seismic_modeling}.)
For more details on the procedure, the reader may refer to Appendix B of
\cite{deheuvels16}. However, in this work, we used the $r_{01}$ rather than the
$r_{010}$ ratio. This allowed us to avoid overfitting and greatly improved the
conditioning of the covariance matrix \citep{Roxburgh2018}.

\subsection{Impact on the seismic observables of a grid of models}
\begin{figure*}
        \includegraphics[width=19cm]{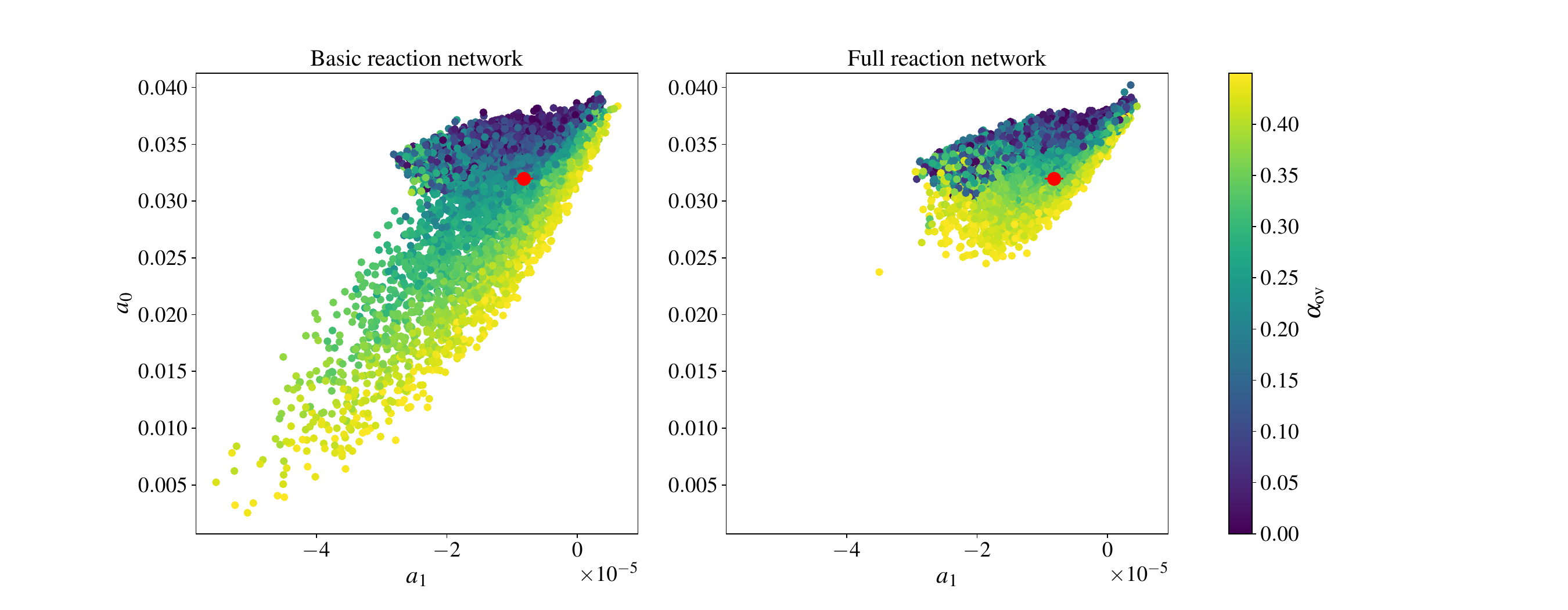}
        \caption{Representation of the models of the grid in the $(a_1, a_0)$ plane. Each point is colored according to the $\ov$ value of the model. The red point represents the observational values for KIC6225718.}
        \label{fig:grid}
\end{figure*}
In this section, we adopt a ``forward'' approach by studying the direct impact
of the change of nuclear network on the seismic observables. To do so, we
computed the $a_0$ and $a_1$ coefficients for all the models of the grid.
Without loss of generality, the $\beta$, $\gamma_1$, and $\gamma_2$ coefficients
were chosen as those obtained from the observed frequencies of KIC6225718 (see
Section~\ref{observational_data}). In order to compare models with equivalent
evolutionary stage, we took models with similar mean density $\bar{\rho}$ by
ensuring that they have similar large separation $\Delta \nu$, as $\Delta \nu
\propto \sqrt{\bar{\rho}}$. More specifically, we ensured that all models have
the same $\nu_{13,0}$ value, namely: $1510\,\mu \mathrm{Hz}$ (the observational
value of KIC6225718). To do so, we linearly interpolated the $a_k$ coefficients
along the track so that the $\nu_{13,0}$ frequency value is reproduced. The
choice of reproducing a low-order frequency rather than $\Delta \nu$ allowed us
to be less sensitive to surface effects, which is necessary when comparing with
observations, as described in Sect.~\ref{seismic_modeling}.

\medskip

Figure~\ref{fig:grid} represents the $a_0$ and $a_1$ values for all the tracks of
the basic network grid (left) or the full network grid (right). Each point is
colored according to the $\ov$ value of the model. For illustration, the red
point indicates (as an example) the observational values for KIC6225718 and their
uncertainties. We can see that modifying the nuclear reaction network has a
strong impact. Indeed, for a given value of $\ov$, models computed with a basic
network have larger convective cores, which leads to lower values of both $a_0$
and $a_1$. Those differences are much greater than the represented uncertainties,
meaning that they are very significant when interpreting observational data. We
can also note that the models which are the most affected are those with
higher values of $\ov$. This is compatible with Eq.~\ref{toy_model_equation},
which indicates that $\Delta r_s$ increases with $\dov$.

We note that even though the largest values of $\ov$ computed here seem high
compared to the ad-hoc value of $0.2$ often quoted in the literature, they are
actually nearly equivalent. Indeed, as mentioned in Section~\ref{carac_model},
the definition of $\dov$ that is used here (i.e., Eq.~\ref{def_ov_mesa}) differs
from the more common one, namely $\dov = \ov \min \left(H_p,
R_{\mathrm{cc}}\right)$. Therefore, a ratio of up to $\conv$ (1.9 in our case) is
found between equivalent overshoot parameter.

\subsection{Impact on seismic modeling}
\label{seismic_modeling}
\subsubsection{Observational data and modeling method}
\label{observational_data}
We seismically modeled two stars observed by \emph{Kepler}, namely, KIC6225718 and
KIC12258514, to quantify the impact of using a basic network on the
inferred stellar parameters. We used the frequencies of \cite{Lund2017} to compute the ratios and the $a_k$ coefficients. To find their uncertainties,
we performed Monte Carlo simulations with 10000 iterations. Finally, we added
two spectroscopic observables: the surface metallicity, $\metal_{\mathrm{spec}}$,
and the effective temperature. $\teff$. Both are taken from \cite{Bruntt2012}.

\bigskip

We found the best models by minimizing the following quantity:
\begin{equation}
        \chi^2 = \sum_{i=1}^{5} \frac{\left(x_i^{\mathrm{mod}} - x_i^{\mathrm{obs}} \right)^2}{\sigma_i^2},
\end{equation}

where $x_i^{\mathrm{obs}}$ are the observables (namely the $a_0$, $a_1$ and
$a_2$ coefficients, $\teff$ and $[Z/X]_{\mathrm{spec}}$), $x_i^{\mathrm{mod}}$
the equivalent values computed with the stellar models, and $\sigma_i$ the
observational uncertainties. We computed the uncertainties on the model
parameters by dividing by 6 the range of parameters of models whose $\chi^2$ is
inferior to $\chi^2_{\min} + 9$.

\subsubsection{Results}
\label{results_seism}

\begin{table*}
        \caption{Parameters of the best model for KIC6225718 and KIC12258514}
        \centering
        \label{table:param_models_6225718}
        
        \begin{tabular}{c c c c c c c c}
                \hline\hline
                Network & $M$ ($\smass$) & $R$ ($R_{\odot}$) & Age (Gyr) &
                $[Z/X]_{\mathrm{init}}$ (dex) & $Y_0$ & $\ov$ & $X_c$ \\
                \hline
                \multicolumn{8}{c}{KIC6225718} \\
                \hline 
                Basic & $1.265 \pm 0.030$ & $1.268 \pm 0.011$ & $1.818 \pm 0.316$ &
                $0.076 \pm 0.064$ & $0.248 \pm 0.012$ & $0.202 \pm 0.044$ & $0.512 \pm
                0.054$  \\
                Full & $1.287 \pm 0.022$ & $1.276 \pm 0.009$ & $1.790 \pm 0.317$ &
                $0.093 \pm 0.055$ & $0.240 \pm 0.011$ & $0.282 \pm 0.052$ & $0.522 \pm
                0.049$   \\
                \hline
                \multicolumn{8}{c}{KIC12258514} \\ 
                \hline 
                Basic & $1.316 \pm 0.017$ & $1.627 \pm 0.009$ & $4.078 \pm 0.156$ &
                $0.150 \pm 0.045$ & $0.251 \pm 0.009$ & $0.124 \pm 0.015$ & $0.060 \pm
                0.012$  \\
                Full & $1.314 \pm 0.027$ & $1.626 \pm 0.012$ & $4.185 \pm 0.256$ &
                $0.121 \pm 0.051$ & $0.243 \pm 0.013$ & $0.154 \pm 0.028$ & $0.058 \pm
                0.012$  \\
                \hline
        \end{tabular}

        \label{parameters_stars}
\end{table*}

The parameters of the best models for the two stars, using either a basic or a
full nuclear reaction network, are summarized in Table~\ref{parameters_stars}.
In both cases, we find that the overshoot parameters of the best models are
significantly different between the two models, with differences approximately
equal to $1.5\,\sigma$. However, all the other stellar parameters are identical
within the uncertainties. We can therefore conclude that (at least for those two
stars) modifying the overshoot parameter alone is enough to compensate the core
size differences caused by the change of nuclear reactions network. 

We note that in this study, we left $\ov$ as a free parameter. However, if it is
fixed, either through a mass-dependent prescription or an ad-hoc value, it may
induce biases on the other parameters of the star.

\section{Discussions and conclusions}
\label{discussions_conclusions}

In this paper, we show how some simplistic assumptions on convection mixing and
nuclear reactions that are commonly made in stellar evolution codes may
impact the size of convective cores of low-mass stars. First, assuming an
instantaneous mixing in the core leads to erroneous central composition
profiles. In particular, lithium has a nuclear timescale (of the order of an hour)
that is much shorter than the convective timescale (of the order of a month): it is
therefore not homogeneous in the core. The resulting difference of composition
between models with a diffusive and an instantaneous mixing impacts the nuclear
production of the pp-2 chain, which can represent up to 20\% of the total
nuclear energy production for $\sim 1.5\,\smass$ stars. Thus, the profile of the
radiative gradient in the core is affected, leading to differences in the
sizes of the convective cores of models with overshoot. Those discrepancies,
which depend mainly on the evolutionary stage, mass, and overshoot
parameter of the model, can represent up to 30\% of the core mass for $\ov =
0.15$. 

Moreover, we observed that using a ``basic'' nuclear reaction network, which
considers beryllium at nuclear equilibrium, has a similar effect. Indeed, the
comparison between the nuclear and convective timescales tells us that beryllium
is actually efficiently mixed while lithium is not. This affects the lithium
composition, which eventually results in core size differences that are similar
to those observed when assuming an instantaneous mixing. Notably, those core
size differences affect the duration of the main-sequence, models with basic
networks having a longer main sequence by around 6\% for $\ov = 0.15$.

We then studied the impact of those core size discrepancies on the seismic
modeling of solar-like oscillators. To do so, we modeled two stars observed by
\emph{Kepler} using ratios of their oscillation frequencies, as those are the
most sensitive to the central structure. We computed models using either a full
nuclear reaction network or a basic one. We found that for the two stars, the
overshoot parameter of the best model with a basic network is significantly
smaller than the overshoot parameter of the best model with a full network.
Apart from that, the other parameters are identical. We conclude that modifying
the overshoot parameter is sufficient to compensate the core size difference
caused by using an overly simplified nuclear reaction network. However, if this
parameter is fixed while modeling the star, those core sizes differences may
induce biases on other parameters. From this result we conclude that it is
necessary, for a proper modeling of low-mass stars with a convective core, to
both consider a diffusive convective mixing and a full nuclear reaction network.
This is particularly important in the framework of the preparation of future
missions such as Plato \citep{plato}, where the proper determination of the age of
main sequence and subgiant stars is crucial. 

\medskip

In this study, we focus on models with step overshooting. However, we note the
interesting result of \cite{Zhang2022}, who pointed out another consequence of the
comparison between $\tau_{\mathrm{conv}}$ and $\tau_{\mathrm{nucl}}$ for models
with exponential overshoot. Indeed, within an exponential overshoot region, the
diffusion mixing coefficient radially varies, and so does $\tau_{\rm conv}$.
Therefore, the radius where $\tau_{\rm conv}$ becomes higher than $\tau_{\rm
nucl}$ (i.e., the radius where the element is not efficiently mixed) is
different for each element. Consequently, the distance over which the elements
are mixed differ depending on their nuclear timescale, as can be seen in the
compositional profiles on Fig.~A1 of that paper. 

Moreover, we only studied the impact on the modeling of solar-like oscillators.
However, there are other types of oscillators found within this mass range.
In particular, $\gamma$ Doradus are stars exhibiting gravity modes during the main
sequence, with masses between approximately $1.5$ and $1.8\,\smass$ (e.g.,
\citealt{Aerts2010}). Their modes are particularly sensitive to the core
properties \citep{Miglio2008} and have already been used to constrain the
overshoot parameter \citep{Mombarg2019,Mombarg2021}. Therefore, the
recommendations of using a full nuclear network and a diffusive mixing should
also be followed when studying this type of oscillator. 

\begin{acknowledgements}
        We thank the anonymous referee for comments that improved the clarity of
        this paper. We also thank Saskia Hekker for constructive comments on earlier
        versions of the draft. We acknowledge support from the Centre National
        d’\'Etudes Spatiales (CNES). A.N. acknowledges funding from the ERC
        Consolidator Grant DipolarSound (grant agreement \#101000296).
\end{acknowledgements}

\bibliographystyle{aa} 
\bibliography{main} 
\end{document}